\documentclass[%
 aip,
 amsmath,amssymb,
 reprint,%
]{revtex4-1}

\usepackage{graphicx}
\usepackage{dcolumn}
\usepackage{bm}

\usepackage[utf8]{inputenc}
\usepackage[T1]{fontenc}
\usepackage{mathptmx}

\usepackage{hyperref}

\newcommand*\conj[1]{
  \vbox{
  \hrule height 0.3ptword
  \kern0.5ex
  \hbox{
  \kern-0.4em
  \ifmmode#1\else\ensuremath{#1}\fi
   \kern-0.em
  }
 }
}
\newcommand{\B}{\boldsymbol}
\newcommand{\tens}[1]{\B{#1}}

\newcommand{\grad}{\B{\mathrm{\nabla}}}
\renewcommand{\div}{\B{\mathrm{\nabla\cdotp}}}

\newcommand{\epsf}{\varepsilon^{\rm f}}
\newcommand{\epsftens}{\tens{\varepsilon}^{\rm f}}
\newcommand{\epstens}{\tens{\varepsilon}}
\newcommand{\epsd}{\varepsilon^{\rm d}}

\newcommand{\epshom}{\tilde{\epstens}}
\newcommand{\fig}[1]{Fig.~(\ref{#1})}
\newcommand{\equ}[1]{Eq.~(\ref{#1})}

\newcommand{\xitens}{\tens{\xi}}
\newcommand{\xihom}{\tilde{\xitens}}

\newcommand{\co}[1]{#1}

\begin{document}

\title{\co{Enhanced tunability in ferroelectric composites through local field enhancement and the effect of disorder}}
\author{Benjamin Vial}
\email{b.vial@qmul.ac.uk}
\author{Yang Hao}
\email{y.hao@qmul.ac.uk}

\affiliation{School of Engineering and Computer Science, Queen Mary, University of London, London, E1 4NS, United Kingdom}
\date{\today}

%
%


%

\begin{abstract}
 We investigate \co{numerically} the homogenized permittivities of \co{composites} made of low index dielectric inclusions in a ferroelectric matrix under
 a static electric field. A refined model is used to take into account the coupling
 between the electrostatic problem and the electric field dependent permittivity of the
 ferroelectric material, leading to a local field enhancement and permittivity change in the ferroelectric.
 Periodic and pseudo-random structures in two dimensions are investigated and
 we \co{compute} the effective permittivity, losses, electrically induced anisotropy and tunability
 of those metamaterials. We show that the tunability of such composites might be substantially
 enhanced in the periodic case, \co{whereas} introducing disorder in the microstructure
 weaken the effect of enhanced local permittivity change. Our results may be useful to guide the synthesis of novel composite ceramics with improved characteristics for controllable microwave devices.
\end{abstract}

\maketitle

 \section{Introduction}
 Ferroelectric materials play a crucial role in reconfigurable
 microwave devices, with typical applications including antenna beam steering,
 phase shifters, tunable power splitters, filters, voltage controlled oscillators and
 matching networks \cite{tagantsev_ferroelectric_2018}. Both bulk ceramics and thin films have
 been employed to design frequency agile components \cite{vendik_ferroelectric_1999,
 lancaster_thin-film_1998,xi_oxide_2000} and metamaterials \cite{hand_frequency_2008, zhao_experimental_2008}.
 The main reason of using
 ferroelectric materials is their strong dependence of their permittivity $\varepsilon$
 on an applied electric field $E$, which is measured by their tunability defined as $n = \varepsilon(0)/\varepsilon(E)$,
 along with a non hysteresis behaviour when used in their paraelectric state.
 The key requirements for antenna and microwave applications are large tunability and low losses.
 These two characteristics are correlated and one has to find a trade-off for optimal
  device performance, which can be quantified by the so called commutation quality factor
 $K = (n -1)^2/(n\, \tan\delta(0)\,\tan\delta(E))$, where $\tan\delta$ is the loss tangent.
 These materials have usually high permittivity values even at microwave frequencies,
 often leading to slow response time and impedance mismatch, which can be an issue in some practical
 applications. Thus it has been considered to mix ferroelectric \co{ceramics with} low-index and
 low-loss non-tunable dielectrics in order to reduce both permittivity values and losses, or to
 use porous ceramics to achieve the same goals without unwanted chemical reactions at the boundaries between dissimilar materials. \co{In particular, the addition of magnesium oxide in barium strontium titanate (BST) ceramics have been shown to decrease the losses while keeping good tunability \cite{irvin_three-dimensional_2005,chung_low-losses_2008}. Ceramics
 such as $\rm Pb(Zr,Ti)O_3$ (PZT) and $\rm BaTiO_3$ (BT) have been used as fillers in polymer
 based composites with high dielectric constant \cite{wang_piezoelectric_2012}. Other mixtures include metal–polymer composites \cite{li_ferroelectric_2013}
 and electroactive polymers such as poly(vinylidene fluoride) (PVDF) with high index dielectric inclusions \cite{hu_preparation_2015}.
 }

The effective parameters of those composites have been investigated
\cite{sherman_ferroelectric-dielectric_2006, jylha_tunability_2008, sherman_tunability_2004, astafiev_can_2003}
and it has been found that the permittivity can be greatly reduced while losses are much less
 sensitive to the dielectric phase addition, and in some situations lead to a small
  increase of the tunability of the mixtures. Analytical models based on the Bruggeman effective
medium approach for low concentration of dielectrics were derived for different configurations (columnar, layered and spherical
inclusions models) and have been successfully compared with numerical simulations and experiments \cite{sherman_ferroelectric-dielectric_2006}.
  In the context of porous ferroelectrics,
   the homogenized properties strongly depends on the size and morphology of the pores
    \cite{okazaki_effects_1973,stanculescu_study_2015}.
    Recently, the concept of tailoring the nonlinear properties of ferroelectric
     and dielectric structures by local field engineering has been introduced \cite{padurariu_tailoring_2012,padurariu_field-dependent_2012,cazacu_tunable_2013}.
      It was shown through finite element calculation including the nonlinear coupling, that by employing
      composite materials made of linear dielectric inclusions into a ferroelectric matrix, one can lower the permittivity while maintaining
      high tunability, due to the local field in the ferroelectric phase which is tuned by the linear dielectric phase.
Moreover, the effect of grain sizes in ferroelectric
ceramics was studied using a model taking the field enhancement into account at
the grain boundaries, and the predicted behaviour successfully compared to experimental data \cite{padurariu_field-dependent_2012}.
Generally, there is a need for refined theoretical and numerical models to explain and design
tunable materials and composites with tailored nonlinear properties.
\co{Note that the general method followed by our coupled model could be applied to other type of tunable
systems where local field enhancement and amplification is relevant, including for example ferromagnetic metamaterials \cite{carignan_ferromagnetic_2011}, liquid crystals based devices \cite{werner_liquid_2007}, or field-enhanced carrier dynamics in doped semiconductors at other frequency ranges, particularly in the THz and near-infrared
\cite{keiser_terahertz_2019, fan_nonlinear_2013}.
}

This study investigates \co{numerically} the effective permittivity of composites made of
 dielectric inclusions in a ferroelectric matrix
by using a two-scale convergence method \cite{allaire_homogenization_1992,
 guenneau_homogenization_2000}.
The originality lies in the fact that a fully coupling model is employed to
calculate the electrostatic field distribution when a uniform biasing field is
applied on the structures, which will result in a local modification of the permittivity
in the ferroelectric phase due to the microstructure. As compared to a simple uncoupled model where the
ferroelectric phase is only modified through the biasing field,
the resulting effective permittivity, dielectric losses, tunability and
anisotropy significantly differ. \co{In contrast with} earlier studies in the
\co{literature} \cite{padurariu_tailoring_2012,padurariu_field-dependent_2012}, we account for
the non-linear coupling beyond the first iteration and use two-scale
convergence homogenization analysis to obtain the effective parameters at
higher frequencies, instead of a capacitance-based model \co{valid in the static regime.
This is an important point, as contrarily to most homogenization procedures that are based on a
quasi-static approximation, the two scale convergence method fixes the frequency and lets the characteristic
size of the system (the periodicity of the composites) tend to zero \cite{guenneau_homogenization_2000}. This asymptotic analysis allows one to study the frequency dependence of the effective parameters.
In addition, analytical models for the effective permittivity routinely employed in the literature
such as Maxwell-Garnett or Bruggeman theories are limited to a few canonical shapes of the inclusions, and
cannot handle arbitrary geometries and media with spatially varying properties. This last point is
of particular importance in the context of this study since we have to account for the field induced local permittivity change.}\\
\co{The model we developed has been implemented with the finite element method (FEM)
and we realise a systematic computational study of ferroelecric-dielectric mixtures.
First, we consider} metamaterials consisting of a square array of parallel
dielectric rods with circular cross section in a ferroelectric host, and
then investigate the effect of random distribution of those rods within the unit cell.


\section{Theory and numerical model}

We consider a composite made of a ferroelectric material with anisotropic
permittivity $\epsftens(\B E)$
that is dependent on an applied electric field $\B E$,
and a non tunable dielectric of permittivity $\epsd$, which are both non-magnetic.
The structures under study are invariant along the $z$ direction, which leads to the standard decomposition
of the wave equation in the transverse electric case (TE, electric field parallel
to the direction of invariance) and the
transverse magnetic case (TM, magnetic field parallel to the direction
of invariance).
A uniform biasing field is applied in order to be able to tune the effective permittivity.
 Modelling homogenized properties of this type of mixtures can be done by assuming
that the electric field distribution is uniform throughout the sample,
so that the study of the tunability is
essentially achieved by changing the value of the properties in the
ferroelectric phase and computing the effective permittivity of the composite.
We refer this approach as to the uncoupled model in the following.
However, a more accurate description is to take into account the change of the
electric field by the microstructure, if any. We therefore need to solve an
electrostatic equation to find the field distribution within the material, but its solution
depends on the permittivities of both materials, and the permittivity in the
ferroelectric phase depends on this induced electric field: this leads to a strongly coupled problem.\\

\subsection{Permittivity model\label{permmodel}}

\begin{figure}[!t]
 \centering
 \includegraphics[width=1\columnwidth]{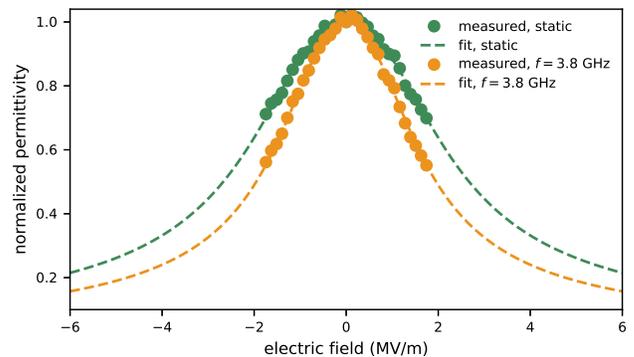}
 \caption{Variation of the ferroelectric permittivity as a function of the
  applied electric field (dots: measurements, dashed lines: fit to
  formula~(\ref{eq_epsf})), for the static case (green) and at microwave frequencies
  (orange, $f=3.8$ GHz). The fitting parameters are given in Table \ref{table_params_fit}.}
 \label{fig1}
\end{figure}
We use barium strontium titanate (BST) as our ferroelectric material.
\co{
${\rm Ba}_{x}{\rm Sr}_{1-x}{\rm TiO}_3$ samples were fabricated using the conventional sintering method with a barium ratio of $x = 0.6$ to obtain a dielectrically tunable material as reported in the literature \cite{agrawal_tunable_2004, hu_preparation_2015}. The tunability was measured using an impedance analyzer up to 100 MHz, and at 3.8 GHz using a loaded microstrip split ring resonator \cite{ansari_design_2015}. The measured tunability of the in-house BST samples of 27\% under 1kv/mm DC bias was in agreement with those reported elsewhere \cite{agrawal_tunable_2004, hu_preparation_2015}. The method presented is however general and only relies of the gradient of the dielectric tunability vs electric field and could be applied to any tunable host material.
The normalized permittivity value
as a function of biasing field are reported on \fig{fig1}.\\}
To describe the permittivity, we make use of the Landau potential
given by $F(P,E) = F_0 +  a P^2/2 + b P^4/4 + cP^6/6 - EP$, where $E$ is
the applied electric field and $P$ is the polarization \cite{landau_electrodynamics_2013, zhou_dielectric_2008}. Variations of the
permittivity with the temperature can be taken into account through the
coefficients $a$, $b$ and $c$, but we assume we are working at a constant
room temperature. We further assume that the material is not subject to any stress, so that the variation
of permittivity due to mechanical constraints is irrelevant.
The equation of state $$\frac{\partial F (P, E)}{\partial P}   = a P_0 + b P_0^3 + c P_0^5 - E = 0$$ gives the
dependence of the polarization on the applied electric field,
with $P_0$ being the equilibrium polarization.
Along the direction of a uniform applied electric field, the relative permittivity is given by:
\begin{equation}
 \epsf(E) = \left[\frac{\partial^2 F (P, E)}{\partial P^2} \right]^{-1} = \frac{\epsf(0)}{1 + \alpha P_0^2 + \beta P_0^4},
 \label{eq_epsf}
\end{equation}
where $\epsf(0)=1/a$, $\alpha=3b/a$ and $\beta=5c/a$. The fitting parameters are given in Table \ref{table_params_fit}.
As the norm of the field increases, the permittivity decreases with a characteristic
bell curve typical for a ferroelectric material in its paraelectric state.
Furthermore, assuming the crystalline principal axes of the ferroelectric material
are oriented in the coordinate directions, and that the diagonal components of the permittivity
tensor are only function of the corresponding bias electric field components \cite{krowne_anisotropic_2002}, we have:
\begin{equation}
 \epsftens (\B E) =
 \left(
 \begin{matrix}
   \epsf_{xx}(E_x) & 0               & 0 \cr
   0               & \epsf_{yy}(E_y) & 0 \cr
   0               & 0               & \epsf_{zz}(E_z) \cr

  \end{matrix}
 \right)
 \label{eq_epsftens}
\end{equation}
where each of the diagonal components have the functional form
given by \equ{eq_epsf}. Note that we will use the static values of permittivity
for the electrostatic modelling, while we are \co{interested} in the homogenized values
of permittivity at microwaves.\\
\begin{table}
 \caption{Fitting parameters to model (\ref{eq_epsf}) for the measured permittivity values as a function
  of applied electric field shown on \fig{fig1}. }
 \label{table_params_fit}
 \begin{ruledtabular}
 \begin{tabular}{llll}
  case        & $\epsf(0)$ & $\alpha$ ($\rm \mu m^2/V^2$) & $\beta$ ($\rm \mu m^4/V^4$) \\
  \hline
  static      & 3050       & 0.120                        & 0.024                       \\
  $f=3.8$ GHz & 165        & 0.240                        & 0.079                       \\
 \end{tabular}
 \end{ruledtabular}
\end{table}

\begin{figure*}[!t]
 \centering
 \includegraphics[width=0.75\textwidth]{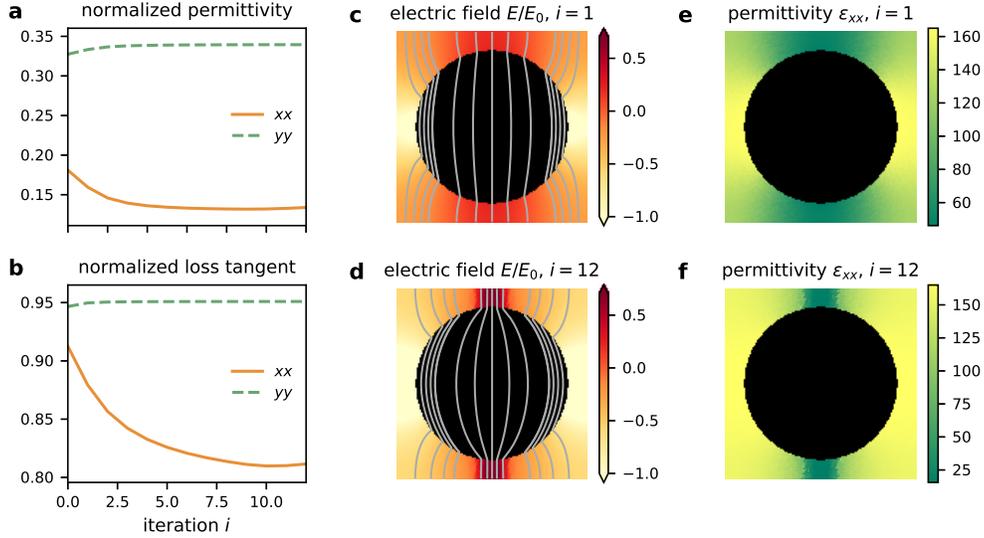}
 \caption{Convergence of the coupled problem.
  Real part (a) and loss tangent (b) of the components of the homogenized
  permittivity tensor as a function of iteration step $i$. \co{The values are normalized to the corresponding quantities for the bulk ferroelectric material.} The distribution of
  the normalized electric field (colour map: magnitude in logarithmic scale,
  lines: equipotential contours) and of the
  $xx$ component of the permittivity tensor are shown for $i=1$
  (c and d) and $i=12$ (e and f).
 }
 \label{conv2D}
\end{figure*}

\subsection{Electrostatic model}
The composites under study are made of two materials, thus their permittivity
is represented by a piecewise defined tensor $\epstens(\B r, \B E)$ which is
equal to $\epsftens(\B E(\B r))$ in the ferroelectric phase and ${\rm diag}(\epsd)$
in the dielectric phase.
In the following, we consider two different cases for the biasing field.
Because of the form~(\ref{eq_epsftens}) assumed for the ferroelectric permittivity
tensor, $\varepsilon_{zz}$ will not be changing for a field in the plane orthogonal
to the $z$ axis. This is the only component
being relevant for TE polarization, so we consider in this case a uniform biasing
electric field applied along the direction of invariance $\B E_0 = E_{0} \B e_z$.
On the other hand,
the in-plane components of $\epsftens$ are tuned by $E_x$ and $E_y$, therefore,
without loss of generality,
we consider a uniform applied electric field directed along the $x$ axis
$\B E_0 = E_{0} \B e_x$ for the TM polarization case.
To calculate the total electric field in the material, \co{one
has to solve for the potential $V$ satisfying Gauss' law:}
\begin{equation}
 \div (\epstens \grad V) = 0
 \label{eq_elstat}
\end{equation}
Note that for the TE case, the solution is trivial since the structures
are invariant along $z$, so that the electric field is equal to the uniform biasing field, and
we will thus not study it in the following.
However in the TM case, the situation is much more complex: this is a coupled problem since the
electric field $\B E=-\grad V$ derived from the
solution of \equ{eq_elstat} depends on the permittivity distribution, which
itself depends on the electric field.
The coupled system formed
of Eqs.~(\ref{eq_epsftens}) and (\ref{eq_elstat}) is solved iteratively until there
is convergence on the norm of the electric field.
Here we would like to emphasise that the permittivity in the ferroelectric material, although
uniform initially, is spatially varying due to the non-uniform distribution
of the total electric field.\\

\subsection{Homogenization}
\co{When the period of the composite metamaterial is much smaller than the wavelength, one can
describe the properties of the composite by a bulk medium with homogenized parameters.}
The effective permittivity for TM polarization is calculated
using a two scale convergence homogenization technique \cite{allaire_homogenization_1992, guenneau_homogenization_2000}.
For this purpose, one has to find the solutions
$\psi_j$ of two annex problems $\mathcal P_j$, $j=\{1, 2\}$:
\begin{equation}
 \div \left[ \xitens \grad(\psi_j + r_j) \right] = 0,
 \label{eq_hom_annex}
\end{equation}
where $\B r = (x, y)^{\rm T}$ is the position vector in the $xy$ plane and
$\xitens=\epstens^{\rm T}/ {\rm det }(\epstens)$.
The homogenized tensor $\xihom$ is obtained with:
\begin{equation}
 \xihom = \langle \xitens \rangle + \B \phi,
 \label{eq_hom}
\end{equation}
where $\langle . \rangle$ denotes the mean value over the unit cell.
The elements of the matrix $\B \phi$ represent correction terms and
are given by $\B \phi_{ij} = \langle \xitens \grad \psi_i \rangle_j$.
Finally the effective permittivity tensor can be calculated using $\epshom=\xihom^{\rm T}/ {\rm det }(\xihom)$.\\
Note that the TE case, which we shall not study here as no coupling happens, is trivial since
the homogenized permittivity is simply the average of the permittivity in the unit cell:
$\epshom = \langle \epstens \rangle$.


\begin{figure*}[!t]
 \centering
 \includegraphics[width=0.75\textwidth]{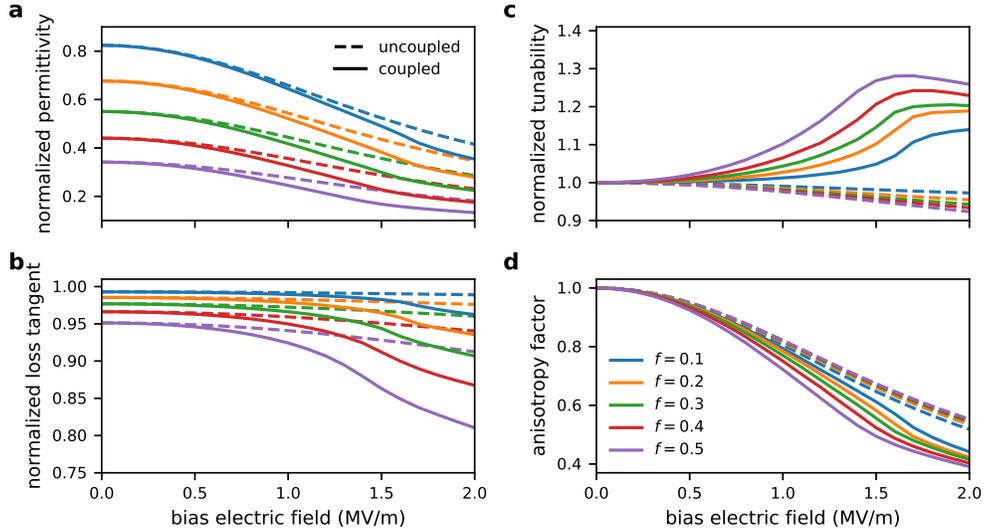}
 \caption{Effective parameters of the 2D metamaterials as a function of the
  applied electric field for various filling fraction of dielectric.
  (a): normalized permittivity, (b): normalized loss tangent, (c): normalized tunability and
  (d): anisotropy factor. The solid lines correspond to the coupled model and
  the dashed lines to the uncoupled model. \co{The values are normalized to the corresponding quantities for the bulk ferroelectric material.}}
 \label{eff_par_2D_TM}
\end{figure*}

\section{Numerical results}
In the following numerical results, the dielectric phase is supposed to be
lossless and non dispersive with $\epsd = 3$ while the ferroelectric material follows the
permittivity described in section~\ref{permmodel} and has a constant loss
tangent $\tan \delta^{\rm f} = 10^{-2}$.
Equations~(\ref{eq_elstat}) and (\ref{eq_hom_annex}) are solved with a Finite Element
Method using the open source packages Gmsh \cite{geuzaine_gmsh_2009} and GetDP \cite{dular_general_1998}.
In both cases we use a square unit cell $\Omega$ of length $d$ with periodic boundary
conditions along $x$ and $y$. Second order Lagrange elements are used and the
solution is computed with a direct solver (MUMPS \cite{amestoy_fully_2001}).

\begin{figure*}[!t]
 \centering
 \includegraphics[width=0.75\textwidth]{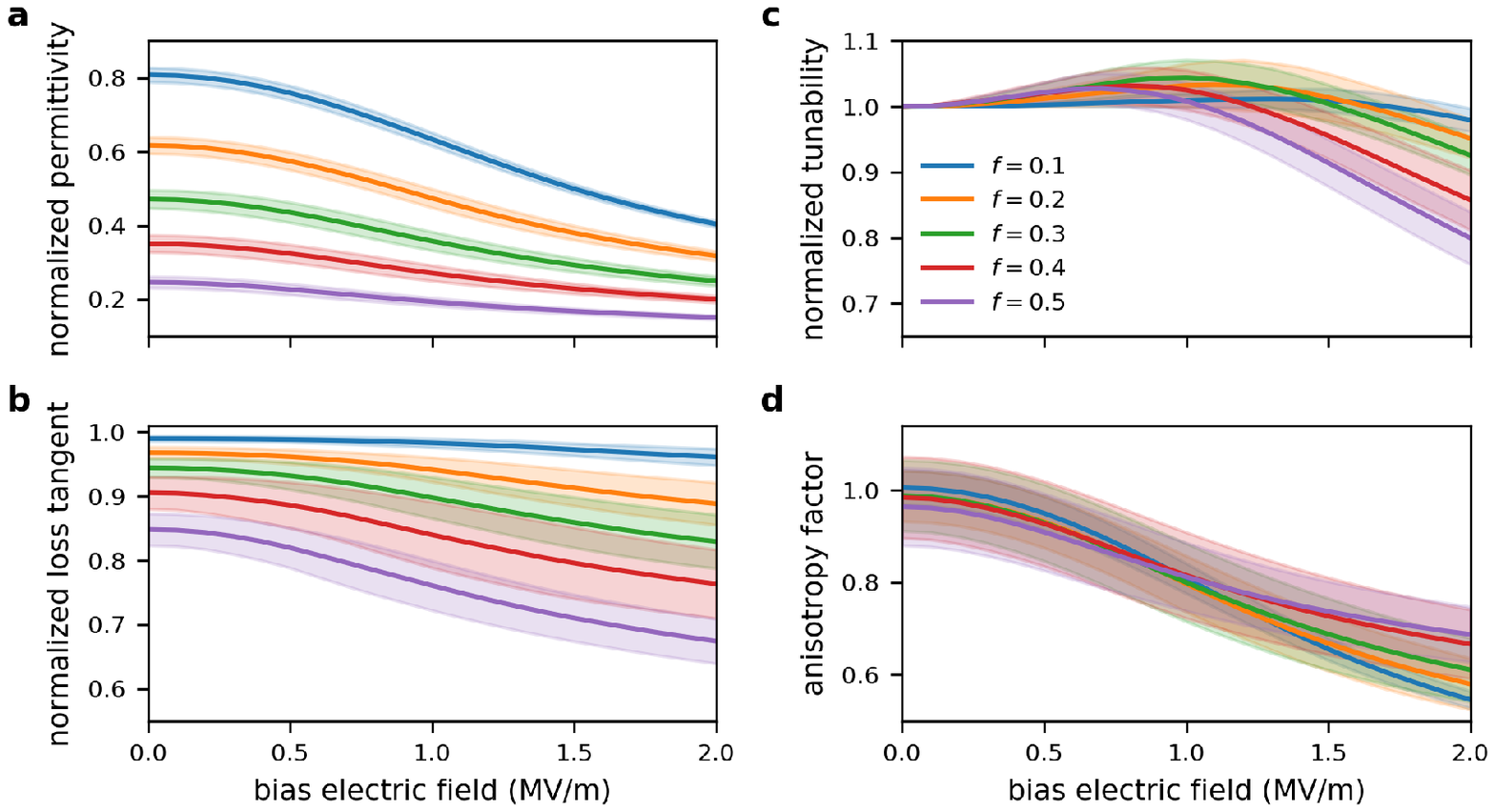}
 \caption{Effective parameters of the pseudo-random composites as a function of the
  applied electric field for various filling fraction of dielectric, when the
  coupling is taken into account.
  (a): normalized permittivity, (b): normalized loss tangent, (c): normalized tunability and
  (d): anisotropy factor. The solid lines represent the average values
  over the 21 samples and the lighter error bands show a confidence interval corresponding to
  the standard deviation. \co{The values are normalized to the corresponding quantities for the bulk ferroelectric material.}}
 \label{eff_par_2Drand_TM}
\end{figure*}
%

\subsection{Two dimensional periodic metamaterial}

Lets us now consider a periodic square array of infinitely long dielectric rods of circular cross section
of radius $r$ embedded in a ferroelectric matrix.\\
We first study the convergence of the coupled problem on the particular case with dielectric
filling fraction $f=\pi r^2/d^2=0.5$ and $E_0=2$MV/m. Figures \ref{conv2D}(a) and \ref{conv2D}(b) show the
convergence of the real part and loss tangent of the components of the homogenized
permittivity tensor, respectively. The $yy$ components converge quickly
and are almost unaffected by the coupling process whereas the
$xx$ components change substantially from the initial conditions.
This is due to the effect of the redistribution
of the electrostatic field within the unit cell (see Figs.~(\ref{conv2D}.c) and (\ref{conv2D}.d)),
where the $x$ component of the electric field is still much stronger
than the $y$ component, even if spatially varying in the ferroelectric medium.
At equilibrium, the electric field is concentrated close to the $y$ axis in between two neighbouring
rods. This in turn affects the permittivity distribution (see Figs.~(\ref{conv2D}.e) and (\ref{conv2D}.f)),
and the homogenized properties of the composite.\\
We computed the effective parameters of these metamaterial structures for different
radii of the rods and studied their behaviour when subjected to an external
electrostatic field (see \fig{eff_par_2D_TM}). The results of our coupled
model differ significantly from the uncoupled one. Increasing the dielectric fraction
lowers the effective permittivity while the losses are slightly reduced but much less sensitive.
Due to the inhomogeneous redistribution of the permittivity over the ferroelectric domain, the
overall tunability changes. In the case studied here, taking into account
the coupling leads to an effective tunability increase with
higher dielectric concentration, and that is larger than the tunability
 of bulk ferroelectric. \co{This can be seen in \fig{eff_par_2D_TM}.c where we plot the tunability of the composites
 along the $x$ axis, $\tilde{n}(E) = \tilde{\varepsilon_{xx}}(E) / \tilde{\varepsilon_{xx}}(0)$, normalized to the tunability of the bulk ferroelectric $n(E) = \epsf_{xx}(E) / \epsf_{xx}(0)$.}
 Two concurrent effects are at stake here: on the one hand
the dilution of ferroelectric makes the composite less tunable, but on the other hand,
the rearrangement of the electrostatic field surrounding the inclusion and its
concentration in some region will cause a higher permittivity change locally.
The relative strength of those phenomena is governed by the shape of the inclusion and its permittivity
and so it is envisioned that the performance of the composites might be enhanced by engineering
their microstructure. \co{Those observations are consistent with previously published numerical and experimental results \cite{padurariu_tailoring_2012}
 where the local field enhancement in porous ferroelectrics has been shown to possibly increase tunability with reducing permittivity for small porosity levels. Our appraoch also agrees with an analytical spherical inclusion model predicting an increase of the tunability with the dilution of the ferroelectric \cite{sherman_ferroelectric-dielectric_2006}.}\\
The geometry of the unit cell is symmetric so the homogenized material is
isotropic when no field is applied.
But when the sample is biased, the permittivity distribution becomes asymmetric due
to the inhomogeneity of the electric field, thus making the effective material properties anisotropic.
This geometric effect is added to the anisotropy arising from the material properties of the ferroelectric
phase itself, and depending on the topology and permittivity of the rods, one effect would be predominant.
In the case studied here, the equilibrium permittivity distribution varies strongly along the bias
 direction and much less orthogonally to it, which adds anisotropy by diminishing the effective
 permittivity in the $x$ direction. This is local field induced effect is what makes the anisotropy stronger
 in our coupled model compared to the uncoupled one (cf. Fig.~(\ref{eff_par_2D_TM}.d) \co{where we plot the anisotropy factor \co{$\alpha =\varepsilon_{xx} / \varepsilon_{yy}$}}).
Those subtle phenomena can only be rigorously taken into account by employing a coupling formalism
and are responsible for the difference observed when compared to a simple uncoupled model.\\
\begin{figure*}[!t]
 \centering
 \includegraphics[width=0.75\textwidth]{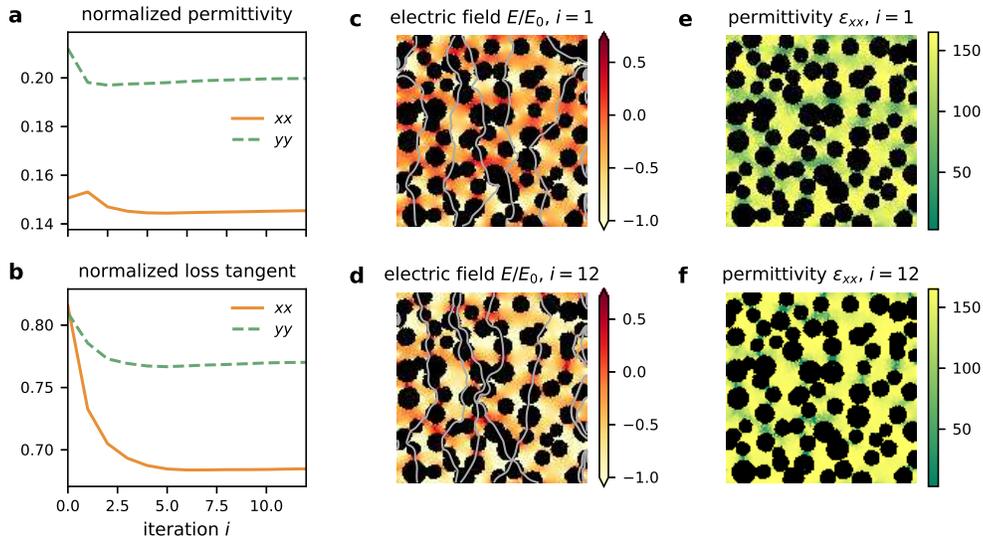}
 \caption{Convergence of the coupled problem in the random case
  for one sample.
  Real part (a) and loss tangent (b) of the components of the homogenized
  permittivity tensor as a function of iteration step $i$ \co{The values are normalized to the corresponding quantities for the bulk ferroelectric material.}. The distribution of
  the normalized electric field (colour map: magnitude in logarithmic scale,
  lines: equipotential contours) and of the $xx$ component of the permittivity tensor are shown for $i=1$
  (c and d) and $i=12$ (e and f).
 }
 \label{conv_random}
\end{figure*}

\subsection{Pseudo-random case}

\begin{figure*}[h!]
 \centering
 \includegraphics[width=0.8\textwidth]{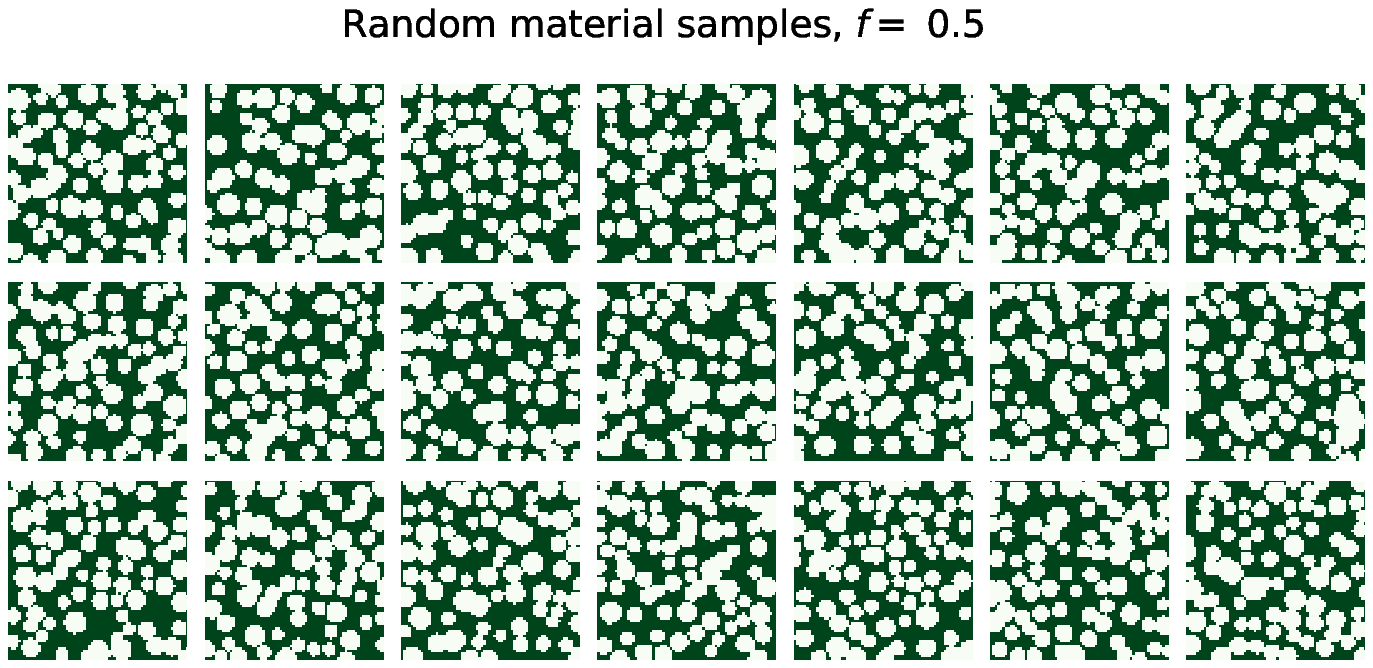}
 \caption{Permittivity distribution of the numerical samples used for $f=0.5$. Dark
  colour indicates the ferroelectric material while light colour represents the
  dielectric inclusions.}
 \label{randmatepsi}
\end{figure*}

\begin{figure*}[h!]
 \centering
 \includegraphics[width=0.8\textwidth]{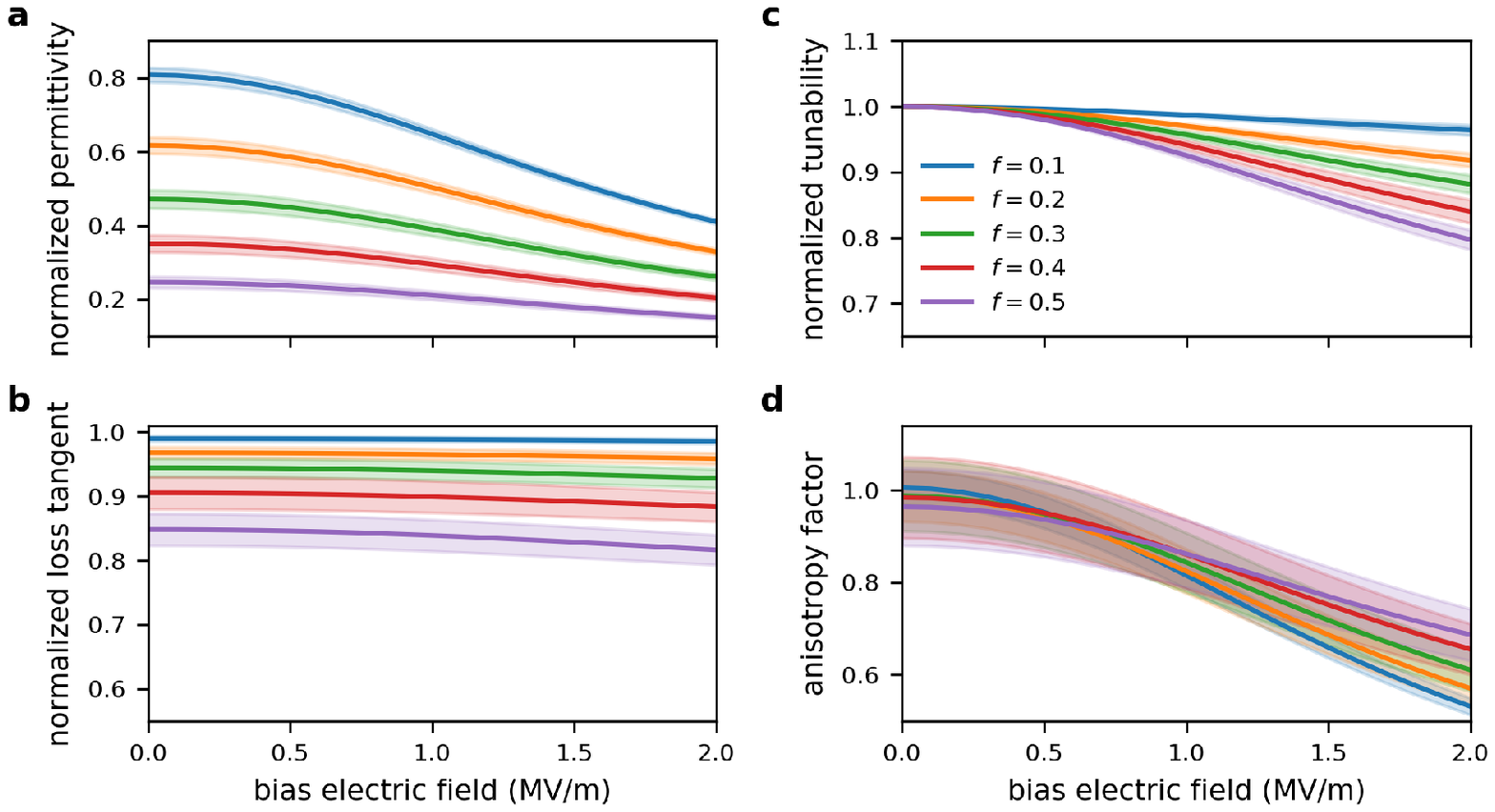}
 \caption{Effective parameters of the random 2D mixtures as a function of the
  applied electric field for various filling fraction of dielectric, when the
  coupling is neglected.
  (a): normalized permittivity, (b): normalized loss tangent, (c): normalized tunability and
  (d): anisotropy factor. The solid lines represent the average values
  over the 21 samples and the lighter error bands show a confidence interval corresponding to
  the standard deviation. \co{The values are normalized to the corresponding quantities for the bulk ferroelectric material.}}
 \label{eff_par_2Drand_TM_uncpl}
\end{figure*}

We finally study the effect of random distribution of the inclusions within the unit cell on the effective parameters of the
composites. This is an important point as fabrication of randomly dispersed
inclusions is much more easy from a technological perspective. For each filling fraction of the dielectric,
we generated 21 numerical samples with inclusions of circular cross section of average radius
$r=d/20$ that can vary by $\pm 30\%$. Their centre is chosen randomly and the
rods are allowed to overlap. An example of distribution for $f=0.5$ is given on \fig{randmatepsi}.
The effective material properties are plotted on \fig{eff_par_2Drand_TM}.
Similarly to the periodic case, the permittivity decreases with increasing dilution of
ferroelectric, but for identical filling fraction,
the permittivity is lower as compared to the periodic array, and the smaller the dielectric concentration the larger
is the difference. Losses decrease as well and the reduction is substantially larger
than the periodic case, with higher variation from sample to sample as $f$ increases.
The effective tunability is on average smaller than that in the periodic case, and
for low biasing fields and for some particular samples can be greater than the bulk tunability. However,
at higher applied electric fields, normalized tunability becomes smaller than unity and
is reduced as one adds more dielectric. For comparison, the homogenized
parameters are plotted on \fig{eff_par_2Drand_TM_uncpl} in the case where the coupling is
neglected. One can see that the coupled and uncoupled models give
similar results for the tunability whereas the losses are still smaller for the coupled case
at higher fields.\\
The redistribution of electric field, permittivity and convergence of
the effective parameters are displayed in \fig{conv_random}. The effect of
disorder plays an important role here: the electrostatic field gets concentrated
in between neighbouring inclusions and the smaller the gap the higher the field, hence
a greater local permittivity change. In addition, even if the distribution is
random, one expects that the anisotropy due to geometry would cancel for a sufficiently large number
of rods (which is the case as the mean anisotropy factor is close to $1$ when no
bias field is applied). However, the anisotropy due to ferroelectric properties is
important in this case as well, as both the $x$ and $y$ components of the electrostatic field
are playing a role.
Because of the relative positions of the rods, both $\varepsilon_{xx}$ and $\varepsilon_{yy}$ are affected
by the coupling, so that the anisotropy factor for higher fields is reduced as compared to the periodic case.
However even if there is a substantial variability from sample to sample, on average, the anisotropy factor
decreases with increasing dielectric concentration.

%

\section{Conclusion}

We have studied the homogenized properties of dielectric/ferroelectric mixtures
using a rigorous model that take into account the coupling between the electrostatic
field distribution and the field dependant ferroelectric permittivity tensor. After
convergence of the coupled problem, the effective permittivity tensor is calculated using
two scale convergence homogenization theory.
The results obtained by this model differ significantly from a simple assumption that
the permittivity of the ferroelectric respond just to the uniform biasing field.
We have considered both periodic and random arrays
of dielectric rods in a ferroelectric matrix in 2D, and studied their effective properties
for TM polarization as a function of dielectric concentration and bias field.
Importantly, adding more low index and low loss dielectric allows to decrease
the overall permittivity significantly and slightly lower the losses.
For the periodic case, the tunability is higher than the bulk due to local field enhancement, whereas
this effect is strongly suppressed when disorder is introduced. The asymmetric redistribution of the permittivity
induce an effective anisotropy that is added to the one arising purely from the ferroelectric material.
The properties of the composites are affected by multiple factors:
geometry and the spatially dependent electric field that will induce locally a tunable, anisotropic
response in the ferroelectric phase depending on its amplitude and direction.
This suggest that the performances of the composites
may be enhanced by distributing the two phases in an optimal way to get high
tunability and low losses. Further work in that direction is needed as well as
extending this study to 3D media.
Finally, because the permittivity of the dielectric is much smaller than the ferroelectric one,
it would be of great interest to use high contrast homogenization theory
\cite{bouchitte_homogenization_2004, cherednichenko_homogenization_2015} to
study this kind of mixtures.
This would reveal the frequency dependant artificial magnetism due to "micro-resonances"
in the high index phase and potentially lead to composites with tunable effective permeability.\\

\begin{acknowledgments}
This work was funded by the Engineering and Physical Sciences Research
Council (EPSRC), UK, under a grant (EP/P005578/1) "Adaptive Tools for
Electromagnetics and Materials Modelling to Bridge the Gap between
Design and Manufacturing (AOTOMAT)".\\
The authors would like to thanks Henry Giddens for performing the
measurements of ferroelectric permittivity used in this paper.\\
%
The codes necessary to reproduce the results in this article are freely
available online at this address: \href{https://www.github.com/benvial/ferromtm}{https://www.github.com/benvial/ferromtm}.
\end{acknowledgments}





\begin{thebibliography}{36}%
\makeatletter
\providecommand \@ifxundefined [1]{%
 \@ifx{#1\undefined}
}%
\providecommand \@ifnum [1]{%
 \ifnum #1\expandafter \@firstoftwo
 \else \expandafter \@secondoftwo
 \fi
}%
\providecommand \@ifx [1]{%
 \ifx #1\expandafter \@firstoftwo
 \else \expandafter \@secondoftwo
 \fi
}%
\providecommand \natexlab [1]{#1}%
\providecommand \enquote  [1]{``#1''}%
\providecommand \bibnamefont  [1]{#1}%
\providecommand \bibfnamefont [1]{#1}%
\providecommand \citenamefont [1]{#1}%
\providecommand \href@noop [0]{\@secondoftwo}%
\providecommand \href [0]{\begingroup \@sanitize@url \@href}%
\providecommand \@href[1]{\@@startlink{#1}\@@href}%
\providecommand \@@href[1]{\endgroup#1\@@endlink}%
\providecommand \@sanitize@url [0]{\catcode `\\12\catcode `\$12\catcode
  `\&12\catcode `\#12\catcode `\^12\catcode `\_12\catcode `\%12\relax}%
\providecommand \@@startlink[1]{}%
\providecommand \@@endlink[0]{}%
\providecommand \url  [0]{\begingroup\@sanitize@url \@url }%
\providecommand \@url [1]{\endgroup\@href {#1}{\urlprefix }}%
\providecommand \urlprefix  [0]{URL }%
\providecommand \Eprint [0]{\href }%
\providecommand \doibase [0]{http://dx.doi.org/}%
\providecommand \selectlanguage [0]{\@gobble}%
\providecommand \bibinfo  [0]{\@secondoftwo}%
\providecommand \bibfield  [0]{\@secondoftwo}%
\providecommand \translation [1]{[#1]}%
\providecommand \BibitemOpen [0]{}%
\providecommand \bibitemStop [0]{}%
\providecommand \bibitemNoStop [0]{.\EOS\space}%
\providecommand \EOS [0]{\spacefactor3000\relax}%
\providecommand \BibitemShut  [1]{\csname bibitem#1\endcsname}%
\let\auto@bib@innerbib\@empty
\bibitem [{\citenamefont {Tagantsev}\ \emph {et~al.}(2018)\citenamefont
  {Tagantsev}, \citenamefont {Sherman}, \citenamefont {Astafiev}, \citenamefont
  {Venkatesh},\ and\ \citenamefont {Setter}}]{tagantsev_ferroelectric_2018}%
  \BibitemOpen
  \bibfield  {author} {\bibinfo {author} {\bibfnamefont {A.~K.}\ \bibnamefont
  {Tagantsev}}, \bibinfo {author} {\bibfnamefont {V.~O.}\ \bibnamefont
  {Sherman}}, \bibinfo {author} {\bibfnamefont {K.~F.}\ \bibnamefont
  {Astafiev}}, \bibinfo {author} {\bibfnamefont {J.}~\bibnamefont {Venkatesh}},
  \ and\ \bibinfo {author} {\bibfnamefont {N.}~\bibnamefont {Setter}},\
  }\bibfield  {title} {\enquote {\bibinfo {title} {Ferroelectric {Materials}
  for {Microwave} {Tunable} {Applications}},}\ }\href@noop {} {\bibfield
  {journal} {\bibinfo  {journal} {J Electroceram}\ }\textbf {\bibinfo {volume}
  {11}},\ \bibinfo {pages} {5--66} (\bibinfo {year} {2018})}\BibitemShut
  {NoStop}%
\bibitem [{\citenamefont {Vendik}\ \emph {et~al.}(1999)\citenamefont {Vendik},
  \citenamefont {Hollmann}, \citenamefont {Kozyrev},\ and\ \citenamefont
  {Prudan}}]{vendik_ferroelectric_1999}%
  \BibitemOpen
  \bibfield  {author} {\bibinfo {author} {\bibfnamefont {O.}~\bibnamefont
  {Vendik}}, \bibinfo {author} {\bibfnamefont {E.}~\bibnamefont {Hollmann}},
  \bibinfo {author} {\bibfnamefont {A.}~\bibnamefont {Kozyrev}}, \ and\
  \bibinfo {author} {\bibfnamefont {A.}~\bibnamefont {Prudan}},\ }\bibfield
  {title} {\enquote {\bibinfo {title} {Ferroelectric tuning of planar and bulk
  microwave devices},}\ }\href@noop {} {\bibfield  {journal} {\bibinfo
  {journal} {J. Supercond.}\ }\textbf {\bibinfo {volume} {12}},\ \bibinfo
  {pages} {325--338} (\bibinfo {year} {1999})}\BibitemShut {NoStop}%
\bibitem [{\citenamefont {Lancaster}, \citenamefont {Powell},\ and\
  \citenamefont {Porch}(1998)}]{lancaster_thin-film_1998}%
  \BibitemOpen
  \bibfield  {author} {\bibinfo {author} {\bibfnamefont {M.}~\bibnamefont
  {Lancaster}}, \bibinfo {author} {\bibfnamefont {J.}~\bibnamefont {Powell}}, \
  and\ \bibinfo {author} {\bibfnamefont {A.}~\bibnamefont {Porch}},\ }\bibfield
   {title} {\enquote {\bibinfo {title} {Thin-film ferroelectric microwave
  devices},}\ }\href@noop {} {\bibfield  {journal} {\bibinfo  {journal}
  {Supercond. Sci. Technol.}\ }\textbf {\bibinfo {volume} {11}},\ \bibinfo
  {pages} {1323} (\bibinfo {year} {1998})}\BibitemShut {NoStop}%
\bibitem [{\citenamefont {Xi}\ \emph {et~al.}(2000)\citenamefont {Xi},
  \citenamefont {Li}, \citenamefont {Si}, \citenamefont {Sirenko},
  \citenamefont {Akimov}, \citenamefont {Fox}, \citenamefont {Clark},\ and\
  \citenamefont {Hao}}]{xi_oxide_2000}%
  \BibitemOpen
  \bibfield  {author} {\bibinfo {author} {\bibfnamefont {X.}~\bibnamefont
  {Xi}}, \bibinfo {author} {\bibfnamefont {H.-C.}\ \bibnamefont {Li}}, \bibinfo
  {author} {\bibfnamefont {W.}~\bibnamefont {Si}}, \bibinfo {author}
  {\bibfnamefont {A.}~\bibnamefont {Sirenko}}, \bibinfo {author} {\bibfnamefont
  {I.}~\bibnamefont {Akimov}}, \bibinfo {author} {\bibfnamefont
  {J.}~\bibnamefont {Fox}}, \bibinfo {author} {\bibfnamefont {A.}~\bibnamefont
  {Clark}}, \ and\ \bibinfo {author} {\bibfnamefont {J.}~\bibnamefont {Hao}},\
  }\bibfield  {title} {\enquote {\bibinfo {title} {Oxide thin films for tunable
  microwave devices},}\ }\href@noop {} {\bibfield  {journal} {\bibinfo
  {journal} {J. Electroceramics}\ }\textbf {\bibinfo {volume} {4}},\ \bibinfo
  {pages} {393--405} (\bibinfo {year} {2000})}\BibitemShut {NoStop}%
\bibitem [{\citenamefont {Hand}\ and\ \citenamefont
  {Cummer}(2008)}]{hand_frequency_2008}%
  \BibitemOpen
  \bibfield  {author} {\bibinfo {author} {\bibfnamefont {T.~H.}\ \bibnamefont
  {Hand}}\ and\ \bibinfo {author} {\bibfnamefont {S.~A.}\ \bibnamefont
  {Cummer}},\ }\bibfield  {title} {\enquote {\bibinfo {title} {Frequency
  tunable electromagnetic metamaterial using ferroelectric loaded split
  rings},}\ }\href@noop {} {\bibfield  {journal} {\bibinfo  {journal} {J. Appl.
  Phys.}\ }\textbf {\bibinfo {volume} {103}},\ \bibinfo {pages} {066105}
  (\bibinfo {year} {2008})}\BibitemShut {NoStop}%
\bibitem [{\citenamefont {Zhao}\ \emph {et~al.}(2008)\citenamefont {Zhao},
  \citenamefont {Kang}, \citenamefont {Zhou}, \citenamefont {Zhao},
  \citenamefont {Li}, \citenamefont {Peng},\ and\ \citenamefont
  {Bai}}]{zhao_experimental_2008}%
  \BibitemOpen
  \bibfield  {author} {\bibinfo {author} {\bibfnamefont {H.}~\bibnamefont
  {Zhao}}, \bibinfo {author} {\bibfnamefont {L.}~\bibnamefont {Kang}}, \bibinfo
  {author} {\bibfnamefont {J.}~\bibnamefont {Zhou}}, \bibinfo {author}
  {\bibfnamefont {Q.}~\bibnamefont {Zhao}}, \bibinfo {author} {\bibfnamefont
  {L.}~\bibnamefont {Li}}, \bibinfo {author} {\bibfnamefont {L.}~\bibnamefont
  {Peng}}, \ and\ \bibinfo {author} {\bibfnamefont {Y.}~\bibnamefont {Bai}},\
  }\bibfield  {title} {\enquote {\bibinfo {title} {Experimental demonstration
  of tunable negative phase velocity and negative refraction in a
  ferromagnetic/ferroelectric composite metamaterial},}\ }\href@noop {}
  {\bibfield  {journal} {\bibinfo  {journal} {Appl. Phys. Lett.}\ }\textbf
  {\bibinfo {volume} {93}},\ \bibinfo {pages} {201106} (\bibinfo {year}
  {2008})}\BibitemShut {NoStop}%
\bibitem [{\citenamefont {Irvin}\ \emph {et~al.}(2005)\citenamefont {Irvin},
  \citenamefont {Levy}, \citenamefont {Guo},\ and\ \citenamefont
  {Bhalla}}]{irvin_three-dimensional_2005}%
  \BibitemOpen
  \bibfield  {author} {\bibinfo {author} {\bibfnamefont {P.}~\bibnamefont
  {Irvin}}, \bibinfo {author} {\bibfnamefont {J.}~\bibnamefont {Levy}},
  \bibinfo {author} {\bibfnamefont {R.}~\bibnamefont {Guo}}, \ and\ \bibinfo
  {author} {\bibfnamefont {A.}~\bibnamefont {Bhalla}},\ }\bibfield  {title}
  {\enquote {\bibinfo {title} {Three-dimensional polarization imaging of
  ({Ba},{Sr}){TiO}3:{MgO} composites},}\ }\href {\doibase 10.1063/1.1854722}
  {\bibfield  {journal} {\bibinfo  {journal} {Appl. Phys. Lett.}\ }\textbf
  {\bibinfo {volume} {86}},\ \bibinfo {pages} {042903} (\bibinfo {year}
  {2005})}\BibitemShut {NoStop}%
\bibitem [{\citenamefont {Chung}\ \emph {et~al.}(2008)\citenamefont {Chung},
  \citenamefont {Elissalde}, \citenamefont {Maglione}, \citenamefont
  {Estournès}, \citenamefont {Paté},\ and\ \citenamefont
  {Ganne}}]{chung_low-losses_2008}%
  \BibitemOpen
  \bibfield  {author} {\bibinfo {author} {\bibfnamefont {U.-C.}\ \bibnamefont
  {Chung}}, \bibinfo {author} {\bibfnamefont {C.}~\bibnamefont {Elissalde}},
  \bibinfo {author} {\bibfnamefont {M.}~\bibnamefont {Maglione}}, \bibinfo
  {author} {\bibfnamefont {C.}~\bibnamefont {Estournès}}, \bibinfo {author}
  {\bibfnamefont {M.}~\bibnamefont {Paté}}, \ and\ \bibinfo {author}
  {\bibfnamefont {J.~P.}\ \bibnamefont {Ganne}},\ }\bibfield  {title} {\enquote
  {\bibinfo {title} {Low-losses, highly tunable {Ba}0.6sr0.4tio3 / {MgO}
  composite},}\ }\href {\doibase 10.1063/1.2837621} {\bibfield  {journal}
  {\bibinfo  {journal} {Appl. Phys. Lett.}\ }\textbf {\bibinfo {volume} {92}},\
  \bibinfo {pages} {042902} (\bibinfo {year} {2008})}\BibitemShut {NoStop}%
\bibitem [{\citenamefont {Wang}\ \emph {et~al.}(2012)\citenamefont {Wang},
  \citenamefont {Luo}, \citenamefont {Ni}, \citenamefont {Du}, \citenamefont
  {Li},\ and\ \citenamefont {Chen}}]{wang_piezoelectric_2012}%
  \BibitemOpen
  \bibfield  {author} {\bibinfo {author} {\bibfnamefont {B.}~\bibnamefont
  {Wang}}, \bibinfo {author} {\bibfnamefont {L.}~\bibnamefont {Luo}}, \bibinfo
  {author} {\bibfnamefont {F.}~\bibnamefont {Ni}}, \bibinfo {author}
  {\bibfnamefont {P.}~\bibnamefont {Du}}, \bibinfo {author} {\bibfnamefont
  {W.}~\bibnamefont {Li}}, \ and\ \bibinfo {author} {\bibfnamefont
  {H.}~\bibnamefont {Chen}},\ }\bibfield  {title} {\enquote {\bibinfo {title}
  {Piezoelectric and ferroelectric properties of ({Bi}1- {xNa}0. 8k0. 2lax) 0.5
  {TiO}3 lead-free ceramics},}\ }\href {\doibase 10.1016/j.jallcom.2012.02.114}
  {\bibfield  {journal} {\bibinfo  {journal} {Journal of Alloys and Compounds}\
  }\textbf {\bibinfo {volume} {526}},\ \bibinfo {pages} {79--84} (\bibinfo
  {year} {2012})}\BibitemShut {NoStop}%
\bibitem [{\citenamefont {Li}\ \emph {et~al.}(2013)\citenamefont {Li},
  \citenamefont {Lim}, \citenamefont {Yao}, \citenamefont {Tay},\ and\
  \citenamefont {Seah}}]{li_ferroelectric_2013}%
  \BibitemOpen
  \bibfield  {author} {\bibinfo {author} {\bibfnamefont {X.}~\bibnamefont
  {Li}}, \bibinfo {author} {\bibfnamefont {Y.-F.}\ \bibnamefont {Lim}},
  \bibinfo {author} {\bibfnamefont {K.}~\bibnamefont {Yao}}, \bibinfo {author}
  {\bibfnamefont {F.~E.~H.}\ \bibnamefont {Tay}}, \ and\ \bibinfo {author}
  {\bibfnamefont {K.~H.}\ \bibnamefont {Seah}},\ }\bibfield  {title} {\enquote
  {\bibinfo {title} {Ferroelectric {Poly}(vinylidene fluoride) {Homopolymer}
  {Nanotubes} {Derived} from {Solution} in {Anodic} {Alumina} {Membrane}
  {Template}},}\ }\href {\doibase 10.1021/cm3028466} {\bibfield  {journal}
  {\bibinfo  {journal} {Chem. Mater.}\ }\textbf {\bibinfo {volume} {25}},\
  \bibinfo {pages} {524--529} (\bibinfo {year} {2013})}\BibitemShut {NoStop}%
\bibitem [{\citenamefont {Hu}\ \emph {et~al.}(2015)\citenamefont {Hu},
  \citenamefont {Gao}, \citenamefont {Kong}, \citenamefont {Yang},
  \citenamefont {Zhang}, \citenamefont {Liu}, \citenamefont {Zhang},\ and\
  \citenamefont {Sun}}]{hu_preparation_2015}%
  \BibitemOpen
  \bibfield  {author} {\bibinfo {author} {\bibfnamefont {G.}~\bibnamefont
  {Hu}}, \bibinfo {author} {\bibfnamefont {F.}~\bibnamefont {Gao}}, \bibinfo
  {author} {\bibfnamefont {J.}~\bibnamefont {Kong}}, \bibinfo {author}
  {\bibfnamefont {S.}~\bibnamefont {Yang}}, \bibinfo {author} {\bibfnamefont
  {Q.}~\bibnamefont {Zhang}}, \bibinfo {author} {\bibfnamefont
  {Z.}~\bibnamefont {Liu}}, \bibinfo {author} {\bibfnamefont {Y.}~\bibnamefont
  {Zhang}}, \ and\ \bibinfo {author} {\bibfnamefont {H.}~\bibnamefont {Sun}},\
  }\bibfield  {title} {\enquote {\bibinfo {title} {Preparation and dielectric
  properties of poly(vinylidene fluoride)/{Ba}0.6sr0.4tio3 composites},}\
  }\href {\doibase 10.1016/j.jallcom.2014.09.005} {\bibfield  {journal}
  {\bibinfo  {journal} {Journal of Alloys and Compounds}\ }\textbf {\bibinfo
  {volume} {619}},\ \bibinfo {pages} {686--692} (\bibinfo {year}
  {2015})}\BibitemShut {NoStop}%
\bibitem [{\citenamefont {Sherman}\ \emph {et~al.}(2006)\citenamefont
  {Sherman}, \citenamefont {Tagantsev}, \citenamefont {Setter}, \citenamefont
  {Iddles},\ and\ \citenamefont
  {Price}}]{sherman_ferroelectric-dielectric_2006}%
  \BibitemOpen
  \bibfield  {author} {\bibinfo {author} {\bibfnamefont {V.~O.}\ \bibnamefont
  {Sherman}}, \bibinfo {author} {\bibfnamefont {A.~K.}\ \bibnamefont
  {Tagantsev}}, \bibinfo {author} {\bibfnamefont {N.}~\bibnamefont {Setter}},
  \bibinfo {author} {\bibfnamefont {D.}~\bibnamefont {Iddles}}, \ and\ \bibinfo
  {author} {\bibfnamefont {T.}~\bibnamefont {Price}},\ }\bibfield  {title}
  {\enquote {\bibinfo {title} {Ferroelectric-dielectric tunable composites},}\
  }\href@noop {} {\bibfield  {journal} {\bibinfo  {journal} {J. Appl. Phys.}\
  }\textbf {\bibinfo {volume} {99}},\ \bibinfo {pages} {074104} (\bibinfo
  {year} {2006})}\BibitemShut {NoStop}%
\bibitem [{\citenamefont {Jylha}\ and\ \citenamefont
  {Sihvola}(2008)}]{jylha_tunability_2008}%
  \BibitemOpen
  \bibfield  {author} {\bibinfo {author} {\bibfnamefont {L.}~\bibnamefont
  {Jylha}}\ and\ \bibinfo {author} {\bibfnamefont {A.~H.}\ \bibnamefont
  {Sihvola}},\ }\bibfield  {title} {\enquote {\bibinfo {title} {Tunability of
  granular ferroelectric dielectric composites},}\ }\href@noop {} {\bibfield
  {journal} {\bibinfo  {journal} {Prog. Electromagn. Res.}\ }\textbf {\bibinfo
  {volume} {78}},\ \bibinfo {pages} {189--207} (\bibinfo {year}
  {2008})}\BibitemShut {NoStop}%
\bibitem [{\citenamefont {Sherman}, \citenamefont {Tagantsev},\ and\
  \citenamefont {Setter}(2004)}]{sherman_tunability_2004}%
  \BibitemOpen
  \bibfield  {author} {\bibinfo {author} {\bibfnamefont {V.~O.}\ \bibnamefont
  {Sherman}}, \bibinfo {author} {\bibfnamefont {A.~K.}\ \bibnamefont
  {Tagantsev}}, \ and\ \bibinfo {author} {\bibfnamefont {N.}~\bibnamefont
  {Setter}},\ }\bibfield  {title} {\enquote {\bibinfo {title} {Tunability and
  loss of the ferroelectric-dielectric composites},}\ }in\ \href@noop {} {\emph
  {\bibinfo {booktitle} {14th {IEEE} {International} {Symposium} on
  {Applications} of {Ferroelectrics}, 2004. {ISAF}-04. 2004}}}\ (\bibinfo
  {year} {2004})\ pp.\ \bibinfo {pages} {33--38}\BibitemShut {NoStop}%
\bibitem [{\citenamefont {Astafiev}\ \emph {et~al.}(2003)\citenamefont
  {Astafiev}, \citenamefont {Sherman}, \citenamefont {Tagantsev},\ and\
  \citenamefont {Setter}}]{astafiev_can_2003}%
  \BibitemOpen
  \bibfield  {author} {\bibinfo {author} {\bibfnamefont {K.~F.}\ \bibnamefont
  {Astafiev}}, \bibinfo {author} {\bibfnamefont {V.~O.}\ \bibnamefont
  {Sherman}}, \bibinfo {author} {\bibfnamefont {A.~K.}\ \bibnamefont
  {Tagantsev}}, \ and\ \bibinfo {author} {\bibfnamefont {N.}~\bibnamefont
  {Setter}},\ }\bibfield  {title} {\enquote {\bibinfo {title} {Can the addition
  of a dielectric improve the figure of merit of a tunable material?}}\
  }\href@noop {} {\bibfield  {journal} {\bibinfo  {journal} {J. Eur. Ceram.
  Soc.}\ }\textbf {\bibinfo {volume} {23}},\ \bibinfo {pages} {2381 -- 2386}
  (\bibinfo {year} {2003})}\BibitemShut {NoStop}%
\bibitem [{\citenamefont {Okazaki}\ and\ \citenamefont
  {Nagata}(1973)}]{okazaki_effects_1973}%
  \BibitemOpen
  \bibfield  {author} {\bibinfo {author} {\bibfnamefont {K.}~\bibnamefont
  {Okazaki}}\ and\ \bibinfo {author} {\bibfnamefont {K.}~\bibnamefont
  {Nagata}},\ }\bibfield  {title} {\enquote {\bibinfo {title} {Effects of
  {Grain} {Size} and {Porosity} on {Electrical} and {Optical} {Properties} of
  {PLZT} {Ceramics}},}\ }\href {\doibase 10.1111/j.1151-2916.1973.tb12363.x}
  {\bibfield  {journal} {\bibinfo  {journal} {J. Am. Ceram. Soc.}\ }\textbf
  {\bibinfo {volume} {56}},\ \bibinfo {pages} {82--86} (\bibinfo {year}
  {1973})}\BibitemShut {NoStop}%
\bibitem [{\citenamefont {Stanculescu}\ \emph {et~al.}(2015)\citenamefont
  {Stanculescu}, \citenamefont {Ciomaga}, \citenamefont {Padurariu},
  \citenamefont {Galizia}, \citenamefont {Horchidan}, \citenamefont {Capiani},
  \citenamefont {Galassi},\ and\ \citenamefont
  {Mitoseriu}}]{stanculescu_study_2015}%
  \BibitemOpen
  \bibfield  {author} {\bibinfo {author} {\bibfnamefont {R.}~\bibnamefont
  {Stanculescu}}, \bibinfo {author} {\bibfnamefont {C.~E.}\ \bibnamefont
  {Ciomaga}}, \bibinfo {author} {\bibfnamefont {L.}~\bibnamefont {Padurariu}},
  \bibinfo {author} {\bibfnamefont {P.}~\bibnamefont {Galizia}}, \bibinfo
  {author} {\bibfnamefont {N.}~\bibnamefont {Horchidan}}, \bibinfo {author}
  {\bibfnamefont {C.}~\bibnamefont {Capiani}}, \bibinfo {author} {\bibfnamefont
  {C.}~\bibnamefont {Galassi}}, \ and\ \bibinfo {author} {\bibfnamefont
  {L.}~\bibnamefont {Mitoseriu}},\ }\bibfield  {title} {\enquote {\bibinfo
  {title} {Study of the role of porosity on the functional properties of
  ({Ba},{Sr}){TiO}3 ceramics},}\ }\href {\doibase
  10.1016/j.jallcom.2015.03.252} {\bibfield  {journal} {\bibinfo  {journal} {J.
  Alloys Compd.}\ }\textbf {\bibinfo {volume} {643}},\ \bibinfo {pages}
  {79--87} (\bibinfo {year} {2015})}\BibitemShut {NoStop}%
\bibitem [{\citenamefont {Padurariu}\ \emph
  {et~al.}(2012{\natexlab{a}})\citenamefont {Padurariu}, \citenamefont
  {Curecheriu}, \citenamefont {Galassi},\ and\ \citenamefont
  {Mitoseriu}}]{padurariu_tailoring_2012}%
  \BibitemOpen
  \bibfield  {author} {\bibinfo {author} {\bibfnamefont {L.}~\bibnamefont
  {Padurariu}}, \bibinfo {author} {\bibfnamefont {L.}~\bibnamefont
  {Curecheriu}}, \bibinfo {author} {\bibfnamefont {C.}~\bibnamefont {Galassi}},
  \ and\ \bibinfo {author} {\bibfnamefont {L.}~\bibnamefont {Mitoseriu}},\
  }\bibfield  {title} {\enquote {\bibinfo {title} {Tailoring non-linear
  dielectric properties by local field engineering in anisotropic porous
  ferroelectric structures},}\ }\href {\doibase 10.1063/1.4729878} {\bibfield
  {journal} {\bibinfo  {journal} {Appl Phys Lett}\ }\textbf {\bibinfo {volume}
  {100}},\ \bibinfo {pages} {252905} (\bibinfo {year}
  {2012}{\natexlab{a}})}\BibitemShut {NoStop}%
\bibitem [{\citenamefont {Padurariu}\ \emph
  {et~al.}(2012{\natexlab{b}})\citenamefont {Padurariu}, \citenamefont
  {Curecheriu}, \citenamefont {Buscaglia},\ and\ \citenamefont
  {Mitoseriu}}]{padurariu_field-dependent_2012}%
  \BibitemOpen
  \bibfield  {author} {\bibinfo {author} {\bibfnamefont {L.}~\bibnamefont
  {Padurariu}}, \bibinfo {author} {\bibfnamefont {L.}~\bibnamefont
  {Curecheriu}}, \bibinfo {author} {\bibfnamefont {V.}~\bibnamefont
  {Buscaglia}}, \ and\ \bibinfo {author} {\bibfnamefont {L.}~\bibnamefont
  {Mitoseriu}},\ }\bibfield  {title} {\enquote {\bibinfo {title}
  {Field-dependent permittivity in nanostructured {BaTiO}\$\{\}\_\{3\}\$
  ceramics: {Modeling} and experimental verification},}\ }\href {\doibase
  10.1103/PhysRevB.85.224111} {\bibfield  {journal} {\bibinfo  {journal} {Phys
  Rev B}\ }\textbf {\bibinfo {volume} {85}},\ \bibinfo {pages} {224111}
  (\bibinfo {year} {2012}{\natexlab{b}})}\BibitemShut {NoStop}%
\bibitem [{\citenamefont {Cazacu}\ \emph {et~al.}(2013)\citenamefont {Cazacu},
  \citenamefont {Curecheriu}, \citenamefont {Neagu}, \citenamefont {Padurariu},
  \citenamefont {Cernescu}, \citenamefont {Lisiecki},\ and\ \citenamefont
  {Mitoseriu}}]{cazacu_tunable_2013}%
  \BibitemOpen
  \bibfield  {author} {\bibinfo {author} {\bibfnamefont {A.}~\bibnamefont
  {Cazacu}}, \bibinfo {author} {\bibfnamefont {L.}~\bibnamefont {Curecheriu}},
  \bibinfo {author} {\bibfnamefont {A.}~\bibnamefont {Neagu}}, \bibinfo
  {author} {\bibfnamefont {L.}~\bibnamefont {Padurariu}}, \bibinfo {author}
  {\bibfnamefont {A.}~\bibnamefont {Cernescu}}, \bibinfo {author}
  {\bibfnamefont {I.}~\bibnamefont {Lisiecki}}, \ and\ \bibinfo {author}
  {\bibfnamefont {L.}~\bibnamefont {Mitoseriu}},\ }\bibfield  {title} {\enquote
  {\bibinfo {title} {Tunable gold-chitosan nanocomposites by local field
  engineering},}\ }\href {\doibase 10.1063/1.4809673} {\bibfield  {journal}
  {\bibinfo  {journal} {Appl Phys Lett}\ }\textbf {\bibinfo {volume} {102}},\
  \bibinfo {pages} {222903} (\bibinfo {year} {2013})}\BibitemShut {NoStop}%
\bibitem [{\citenamefont {Carignan}\ \emph {et~al.}(2011)\citenamefont
  {Carignan}, \citenamefont {Yelon}, \citenamefont {Menard},\ and\
  \citenamefont {Caloz}}]{carignan_ferromagnetic_2011}%
  \BibitemOpen
  \bibfield  {author} {\bibinfo {author} {\bibfnamefont {L.}~\bibnamefont
  {Carignan}}, \bibinfo {author} {\bibfnamefont {A.}~\bibnamefont {Yelon}},
  \bibinfo {author} {\bibfnamefont {D.}~\bibnamefont {Menard}}, \ and\ \bibinfo
  {author} {\bibfnamefont {C.}~\bibnamefont {Caloz}},\ }\bibfield  {title}
  {\enquote {\bibinfo {title} {Ferromagnetic {Nanowire} {Metamaterials}:
  {Theory} and {Applications}},}\ }\href {\doibase 10.1109/TMTT.2011.2163202}
  {\bibfield  {journal} {\bibinfo  {journal} {IEEE Trans. Microw. Theory
  Tech.}\ }\textbf {\bibinfo {volume} {59}},\ \bibinfo {pages} {2568--2586}
  (\bibinfo {year} {2011})}\BibitemShut {NoStop}%
\bibitem [{\citenamefont {Werner}\ \emph {et~al.}(2007)\citenamefont {Werner},
  \citenamefont {Kwon}, \citenamefont {Khoo}, \citenamefont {Kildishev},\ and\
  \citenamefont {Shalaev}}]{werner_liquid_2007}%
  \BibitemOpen
  \bibfield  {author} {\bibinfo {author} {\bibfnamefont {D.~H.}\ \bibnamefont
  {Werner}}, \bibinfo {author} {\bibfnamefont {D.-H.}\ \bibnamefont {Kwon}},
  \bibinfo {author} {\bibfnamefont {I.-C.}\ \bibnamefont {Khoo}}, \bibinfo
  {author} {\bibfnamefont {A.~V.}\ \bibnamefont {Kildishev}}, \ and\ \bibinfo
  {author} {\bibfnamefont {V.~M.}\ \bibnamefont {Shalaev}},\ }\bibfield
  {title} {\enquote {\bibinfo {title} {Liquid crystal clad near-infrared
  metamaterials with tunable negative-zero-positive refractive indices},}\
  }\href {\doibase 10.1364/OE.15.003342} {\bibfield  {journal} {\bibinfo
  {journal} {Opt. Express, OE}\ }\textbf {\bibinfo {volume} {15}},\ \bibinfo
  {pages} {3342--3347} (\bibinfo {year} {2007})}\BibitemShut {NoStop}%
\bibitem [{\citenamefont {Keiser}\ and\ \citenamefont
  {Klarskov}(2019)}]{keiser_terahertz_2019}%
  \BibitemOpen
  \bibfield  {author} {\bibinfo {author} {\bibfnamefont {G.~R.}\ \bibnamefont
  {Keiser}}\ and\ \bibinfo {author} {\bibfnamefont {P.}~\bibnamefont
  {Klarskov}},\ }\bibfield  {title} {\enquote {\bibinfo {title} {Terahertz
  {Field} {Confinement} in {Nonlinear} {Metamaterials} and {Near}-{Field}
  {Imaging}},}\ }\href {\doibase 10.3390/photonics6010022} {\bibfield
  {journal} {\bibinfo  {journal} {Photonics}\ }\textbf {\bibinfo {volume}
  {6}},\ \bibinfo {pages} {22} (\bibinfo {year} {2019})}\BibitemShut {NoStop}%
\bibitem [{\citenamefont {Fan}\ \emph {et~al.}(2013)\citenamefont {Fan},
  \citenamefont {Hwang}, \citenamefont {Liu}, \citenamefont {Strikwerda},
  \citenamefont {Sternbach}, \citenamefont {Zhang}, \citenamefont {Zhao},
  \citenamefont {Zhang}, \citenamefont {Nelson},\ and\ \citenamefont
  {Averitt}}]{fan_nonlinear_2013}%
  \BibitemOpen
  \bibfield  {author} {\bibinfo {author} {\bibfnamefont {K.}~\bibnamefont
  {Fan}}, \bibinfo {author} {\bibfnamefont {H.~Y.}\ \bibnamefont {Hwang}},
  \bibinfo {author} {\bibfnamefont {M.}~\bibnamefont {Liu}}, \bibinfo {author}
  {\bibfnamefont {A.~C.}\ \bibnamefont {Strikwerda}}, \bibinfo {author}
  {\bibfnamefont {A.}~\bibnamefont {Sternbach}}, \bibinfo {author}
  {\bibfnamefont {J.}~\bibnamefont {Zhang}}, \bibinfo {author} {\bibfnamefont
  {X.}~\bibnamefont {Zhao}}, \bibinfo {author} {\bibfnamefont {X.}~\bibnamefont
  {Zhang}}, \bibinfo {author} {\bibfnamefont {K.~A.}\ \bibnamefont {Nelson}}, \
  and\ \bibinfo {author} {\bibfnamefont {R.~D.}\ \bibnamefont {Averitt}},\
  }\bibfield  {title} {\enquote {\bibinfo {title} {Nonlinear {Terahertz}
  {Metamaterials} via {Field}-{Enhanced} {Carrier} {Dynamics} in {GaAs}},}\
  }\href {\doibase 10.1103/PhysRevLett.110.217404} {\bibfield  {journal}
  {\bibinfo  {journal} {Phys. Rev. Lett.}\ }\textbf {\bibinfo {volume} {110}},\
  \bibinfo {pages} {217404} (\bibinfo {year} {2013})}\BibitemShut {NoStop}%
\bibitem [{\citenamefont {Allaire}(1992)}]{allaire_homogenization_1992}%
  \BibitemOpen
  \bibfield  {author} {\bibinfo {author} {\bibfnamefont {G.}~\bibnamefont
  {Allaire}},\ }\bibfield  {title} {\enquote {\bibinfo {title} {Homogenization
  and {Two}-{Scale} {Convergence}},}\ }\href@noop {} {\bibfield  {journal}
  {\bibinfo  {journal} {SIAM J. Math. Anal.}\ }\textbf {\bibinfo {volume}
  {23}},\ \bibinfo {pages} {1482--1518} (\bibinfo {year} {1992})}\BibitemShut
  {NoStop}%
\bibitem [{\citenamefont {Guenneau}\ and\ \citenamefont
  {Zolla}(2000)}]{guenneau_homogenization_2000}%
  \BibitemOpen
  \bibfield  {author} {\bibinfo {author} {\bibfnamefont {S.}~\bibnamefont
  {Guenneau}}\ and\ \bibinfo {author} {\bibfnamefont {F.}~\bibnamefont
  {Zolla}},\ }\bibfield  {title} {\enquote {\bibinfo {title} {Homogenization of
  {Three}-{Dimensional} {Finite} {Photonic} {Crystals}},}\ }\href@noop {}
  {\bibfield  {journal} {\bibinfo  {journal} {J. Electromagn. Waves Appl.}\
  }\textbf {\bibinfo {volume} {14}},\ \bibinfo {pages} {529--530} (\bibinfo
  {year} {2000})}\BibitemShut {NoStop}%
\bibitem [{\citenamefont {Agrawal}\ \emph {et~al.}(2004)\citenamefont
  {Agrawal}, \citenamefont {Guo}, \citenamefont {Agrawal},\ and\ \citenamefont
  {Bhalla}}]{agrawal_tunable_2004}%
  \BibitemOpen
  \bibfield  {author} {\bibinfo {author} {\bibfnamefont {S.}~\bibnamefont
  {Agrawal}}, \bibinfo {author} {\bibfnamefont {R.}~\bibnamefont {Guo}},
  \bibinfo {author} {\bibfnamefont {D.}~\bibnamefont {Agrawal}}, \ and\
  \bibinfo {author} {\bibfnamefont {A.~S.}\ \bibnamefont {Bhalla}},\ }\bibfield
   {title} {\enquote {\bibinfo {title} {Tunable {BST}:{MgO} {Dielectric}
  {Composite} by {Microwave} {Sintering}},}\ }\href {\doibase
  10.1080/00150190490460803} {\bibfield  {journal} {\bibinfo  {journal}
  {Ferroelectrics}\ }\textbf {\bibinfo {volume} {306}},\ \bibinfo {pages}
  {155--163} (\bibinfo {year} {2004})}\BibitemShut {NoStop}%
\bibitem [{\citenamefont {Ansari}, \citenamefont {Jha},\ and\ \citenamefont
  {Akhtar}(2015)}]{ansari_design_2015}%
  \BibitemOpen
  \bibfield  {author} {\bibinfo {author} {\bibfnamefont {M.~A.~H.}\
  \bibnamefont {Ansari}}, \bibinfo {author} {\bibfnamefont {A.~K.}\
  \bibnamefont {Jha}}, \ and\ \bibinfo {author} {\bibfnamefont {M.~J.}\
  \bibnamefont {Akhtar}},\ }\bibfield  {title} {\enquote {\bibinfo {title}
  {Design and {Application} of the {CSRR}-{Based} {Planar} {Sensor} for
  {Noninvasive} {Measurement} of {Complex} {Permittivity}},}\ }\href {\doibase
  10.1109/JSEN.2015.2469683} {\bibfield  {journal} {\bibinfo  {journal} {IEEE
  Sens. J.}\ }\textbf {\bibinfo {volume} {15}},\ \bibinfo {pages} {7181--7189}
  (\bibinfo {year} {2015})}\BibitemShut {NoStop}%
\bibitem [{\citenamefont {Landau}\ \emph {et~al.}(2013)\citenamefont {Landau},
  \citenamefont {Bell}, \citenamefont {Kearsley}, \citenamefont {Pitaevskii},
  \citenamefont {Lifshitz},\ and\ \citenamefont
  {Sykes}}]{landau_electrodynamics_2013}%
  \BibitemOpen
  \bibfield  {author} {\bibinfo {author} {\bibfnamefont {L.~D.}\ \bibnamefont
  {Landau}}, \bibinfo {author} {\bibfnamefont {J.}~\bibnamefont {Bell}},
  \bibinfo {author} {\bibfnamefont {M.}~\bibnamefont {Kearsley}}, \bibinfo
  {author} {\bibfnamefont {L.}~\bibnamefont {Pitaevskii}}, \bibinfo {author}
  {\bibfnamefont {E.}~\bibnamefont {Lifshitz}}, \ and\ \bibinfo {author}
  {\bibfnamefont {J.}~\bibnamefont {Sykes}},\ }\href@noop {} {\emph {\bibinfo
  {title} {Electrodynamics of continuous media}}},\ Vol.~\bibinfo {volume} {8}\
  (\bibinfo  {publisher} {elsevier},\ \bibinfo {year} {2013})\BibitemShut
  {NoStop}%
\bibitem [{\citenamefont {Zhou}\ \emph {et~al.}(2008)\citenamefont {Zhou},
  \citenamefont {Boggs}, \citenamefont {Ramprasad}, \citenamefont {Aindow},
  \citenamefont {Erkey},\ and\ \citenamefont {Alpay}}]{zhou_dielectric_2008}%
  \BibitemOpen
  \bibfield  {author} {\bibinfo {author} {\bibfnamefont {K.}~\bibnamefont
  {Zhou}}, \bibinfo {author} {\bibfnamefont {S.~A.}\ \bibnamefont {Boggs}},
  \bibinfo {author} {\bibfnamefont {R.}~\bibnamefont {Ramprasad}}, \bibinfo
  {author} {\bibfnamefont {M.}~\bibnamefont {Aindow}}, \bibinfo {author}
  {\bibfnamefont {C.}~\bibnamefont {Erkey}}, \ and\ \bibinfo {author}
  {\bibfnamefont {S.~P.}\ \bibnamefont {Alpay}},\ }\bibfield  {title} {\enquote
  {\bibinfo {title} {Dielectric response and tunability of a
  dielectric-paraelectric composite},}\ }\href@noop {} {\bibfield  {journal}
  {\bibinfo  {journal} {Appl. Phys. Lett.}\ }\textbf {\bibinfo {volume} {93}},\
  \bibinfo {pages} {102908} (\bibinfo {year} {2008})}\BibitemShut {NoStop}%
\bibitem [{\citenamefont {Krowne}\ \emph {et~al.}(2002)\citenamefont {Krowne},
  \citenamefont {Daniel}, \citenamefont {Kirchoefer},\ and\ \citenamefont
  {Pond}}]{krowne_anisotropic_2002}%
  \BibitemOpen
  \bibfield  {author} {\bibinfo {author} {\bibfnamefont {C.~M.}\ \bibnamefont
  {Krowne}}, \bibinfo {author} {\bibfnamefont {M.}~\bibnamefont {Daniel}},
  \bibinfo {author} {\bibfnamefont {S.~W.}\ \bibnamefont {Kirchoefer}}, \ and\
  \bibinfo {author} {\bibfnamefont {J.~A.}\ \bibnamefont {Pond}},\ }\bibfield
  {title} {\enquote {\bibinfo {title} {Anisotropic permittivity and attenuation
  extraction from propagation constant measurements using an anisotropic
  full-wave {Green}'s function solver for coplanar ferroelectric thin-film
  devices},}\ }\href@noop {} {\bibfield  {journal} {\bibinfo  {journal} {IEEE
  Trans. Microw. Theory Tech.}\ }\textbf {\bibinfo {volume} {50}},\ \bibinfo
  {pages} {537--548} (\bibinfo {year} {2002})}\BibitemShut {NoStop}%
\bibitem [{\citenamefont {Geuzaine}\ and\ \citenamefont
  {Remacle}(2009)}]{geuzaine_gmsh_2009}%
  \BibitemOpen
  \bibfield  {author} {\bibinfo {author} {\bibfnamefont {C.}~\bibnamefont
  {Geuzaine}}\ and\ \bibinfo {author} {\bibfnamefont {J.-F.}\ \bibnamefont
  {Remacle}},\ }\bibfield  {title} {\enquote {\bibinfo {title} {Gmsh: {A} 3-{D}
  finite element mesh generator with built-in pre- and post-processing
  facilities},}\ }\href {\doibase 10.1002/nme.2579} {\bibfield  {journal}
  {\bibinfo  {journal} {Int. J. Numer. Methods Eng.}\ }\textbf {\bibinfo
  {volume} {79}},\ \bibinfo {pages} {1309--1331} (\bibinfo {year}
  {2009})}\BibitemShut {NoStop}%
\bibitem [{\citenamefont {Dular}\ \emph {et~al.}(1998)\citenamefont {Dular},
  \citenamefont {Geuzaine}, \citenamefont {Henrotte},\ and\ \citenamefont
  {Legros}}]{dular_general_1998}%
  \BibitemOpen
  \bibfield  {author} {\bibinfo {author} {\bibfnamefont {P.}~\bibnamefont
  {Dular}}, \bibinfo {author} {\bibfnamefont {C.}~\bibnamefont {Geuzaine}},
  \bibinfo {author} {\bibfnamefont {F.}~\bibnamefont {Henrotte}}, \ and\
  \bibinfo {author} {\bibfnamefont {W.}~\bibnamefont {Legros}},\ }\bibfield
  {title} {\enquote {\bibinfo {title} {A general environment for the treatment
  of discrete problems and its application to the finite element method},}\
  }\href@noop {} {\bibfield  {journal} {\bibinfo  {journal} {IEEE Trans.
  Magn.}\ }\textbf {\bibinfo {volume} {34}},\ \bibinfo {pages} {3395--3398}
  (\bibinfo {year} {1998})}\BibitemShut {NoStop}%
\bibitem [{\citenamefont {Amestoy}\ \emph {et~al.}(2001)\citenamefont
  {Amestoy}, \citenamefont {Duff}, \citenamefont {Koster},\ and\ \citenamefont
  {L'Excellent}}]{amestoy_fully_2001}%
  \BibitemOpen
  \bibfield  {author} {\bibinfo {author} {\bibfnamefont {P.~R.}\ \bibnamefont
  {Amestoy}}, \bibinfo {author} {\bibfnamefont {I.~S.}\ \bibnamefont {Duff}},
  \bibinfo {author} {\bibfnamefont {J.}~\bibnamefont {Koster}}, \ and\ \bibinfo
  {author} {\bibfnamefont {J.-Y.}\ \bibnamefont {L'Excellent}},\ }\bibfield
  {title} {\enquote {\bibinfo {title} {A {Fully} {Asynchronous} {Multifrontal}
  {Solver} {Using} {Distributed} {Dynamic} {Scheduling}},}\ }\href@noop {}
  {\bibfield  {journal} {\bibinfo  {journal} {SIAM J. Matrix Anal. Appl.}\
  }\textbf {\bibinfo {volume} {23}},\ \bibinfo {pages} {15--41} (\bibinfo
  {year} {2001})}\BibitemShut {NoStop}%
\bibitem [{\citenamefont {Bouchitté}\ and\ \citenamefont
  {Felbacq}(2004)}]{bouchitte_homogenization_2004}%
  \BibitemOpen
  \bibfield  {author} {\bibinfo {author} {\bibfnamefont {G.}~\bibnamefont
  {Bouchitté}}\ and\ \bibinfo {author} {\bibfnamefont {D.}~\bibnamefont
  {Felbacq}},\ }\bibfield  {title} {\enquote {\bibinfo {title} {Homogenization
  near resonances and artificial magnetism from dielectrics},}\ }\href@noop {}
  {\bibfield  {journal} {\bibinfo  {journal} {Comptes Rendus Mathématique}\
  }\textbf {\bibinfo {volume} {339}},\ \bibinfo {pages} {377 -- 382} (\bibinfo
  {year} {2004})}\BibitemShut {NoStop}%
\bibitem [{\citenamefont {Cherednichenko}\ and\ \citenamefont
  {Cooper}(2015)}]{cherednichenko_homogenization_2015}%
  \BibitemOpen
  \bibfield  {author} {\bibinfo {author} {\bibfnamefont {K.}~\bibnamefont
  {Cherednichenko}}\ and\ \bibinfo {author} {\bibfnamefont {S.}~\bibnamefont
  {Cooper}},\ }\bibfield  {title} {\enquote {\bibinfo {title} {Homogenization
  of the system of high-contrast {Maxwell} equations},}\ }\href@noop {}
  {\bibfield  {journal} {\bibinfo  {journal} {Mathematika}\ }\textbf {\bibinfo
  {volume} {61}},\ \bibinfo {pages} {475--500} (\bibinfo {year}
  {2015})}\BibitemShut {NoStop}%
\end{thebibliography}

%

\end{document}